\documentclass[twocolumn,prl,showpacs]{revtex4}
\usepackage{graphicx,enumerate}
\usepackage{amsmath}
\usepackage{amssymb}
\usepackage{subfigure}
\usepackage{verbatim} 
\usepackage{float}

\newcommand{\gl}{\mathrel{\raise0.6ex\hbox{$>$\kern-.75em\lower1ex\hbox{$<$}}}}
\begin{document}
\title{ Fate of 2D Kinetic Ferromagnets and Critical Percolation
  Crossing Probabilities}

\author{J. Olejarz}
\author{P. L. Krapivsky}
\author{S. Redner}
\affiliation{Center for Polymer Studies and Department of Physics, Boston University, Boston, MA 02215}

\begin{abstract}
  We present evidence for a deep connection between the zero-temperature
  coarsening of both the two-dimensional time-dependent Ginzburg-Landau
  (TDGL) equation and the kinetic Ising model (KIM) with critical continuum
  percolation.  In addition to reaching the ground state, the TDGL and KIM
  can fall into a variety of topologically distinct metastable stripe states.
  The probability to reach a stripe state that winds $a$ times horizontally
  and $b$ times vertically on a square lattice with periodic boundary
  conditions equals the corresponding exactly-solved critical percolation
  crossing probability $\mathcal{P}_{a,b}$ for a spanning path with winding
  numbers $a$ and $b$.
\end{abstract}
\pacs{64.60.My, 05.40.-a, 05.50.+q, 75.40.Gb}
\maketitle

When a ferromagnet with non-conserved spin flip dynamics is quenched from
above the critical temperature to zero temperature, a beautiful coarsening
domain mosaic emerges~\cite{lif_62,gunton_dynamics_1983,bray_review}
(Fig.~1).  For finite systems, this coarsening ends when the typical domain
length reaches the linear dimension of the system.  What is the resulting
final state?  A naive expectation is that the ground state is ultimately
reached because each microscopic spin update either decreases or maintains
the energy of the system.  However, this lowest-energy state is not
necessarily the final outcome.  There exist a plethora of metastable states,
such as straight stripes in two dimensions~\cite{spirin_fate_2001,ONSS,KS}
and more bizarre gyroid or ``plumber's nightmare'' states in three
dimensions~\cite{OKR}, which are infinitely long lived at zero temperature.
Once the system falls into such a state, the only escape route is via
energy-raising spin flips.  Since such events do not occur at zero
temperature, there is no escape to the ground state.

In the intermediate-time regime, where the typical domain size substantially
exceeds the lattice spacing but is much smaller than the system size, the
domain mosaic visually resembles the cluster geometry of continuum
percolation~\cite{BKR09}.  This correspondence has sparked recent work on
possible connections between these seemingly disparate
models~\cite{BKR09,arenzon_exact_2007}.  In two dimensions, continuum
percolation is critical when the concentrations of both phases are
equal~\cite{continuum}.  This duality explains why the ground state
corresponding to the majority phase is always reached in coarsening in the
thermodynamic limit for non-zero initial
magnetization~\cite{spirin_fate_2001}.  In this case, the majority phase
percolates in all directions and inevitably engulfs the entire system.  The
most interesting situation of quenching from above the critical temperature
corresponds to zero initial magnetization, so that the system in the
intermediate-time regime is at the critical point of two-dimensional
continuum percolation.

\begin{figure}[ht]
\label{snapshots}
\subfigure[]{\includegraphics[width=0.116\textwidth]{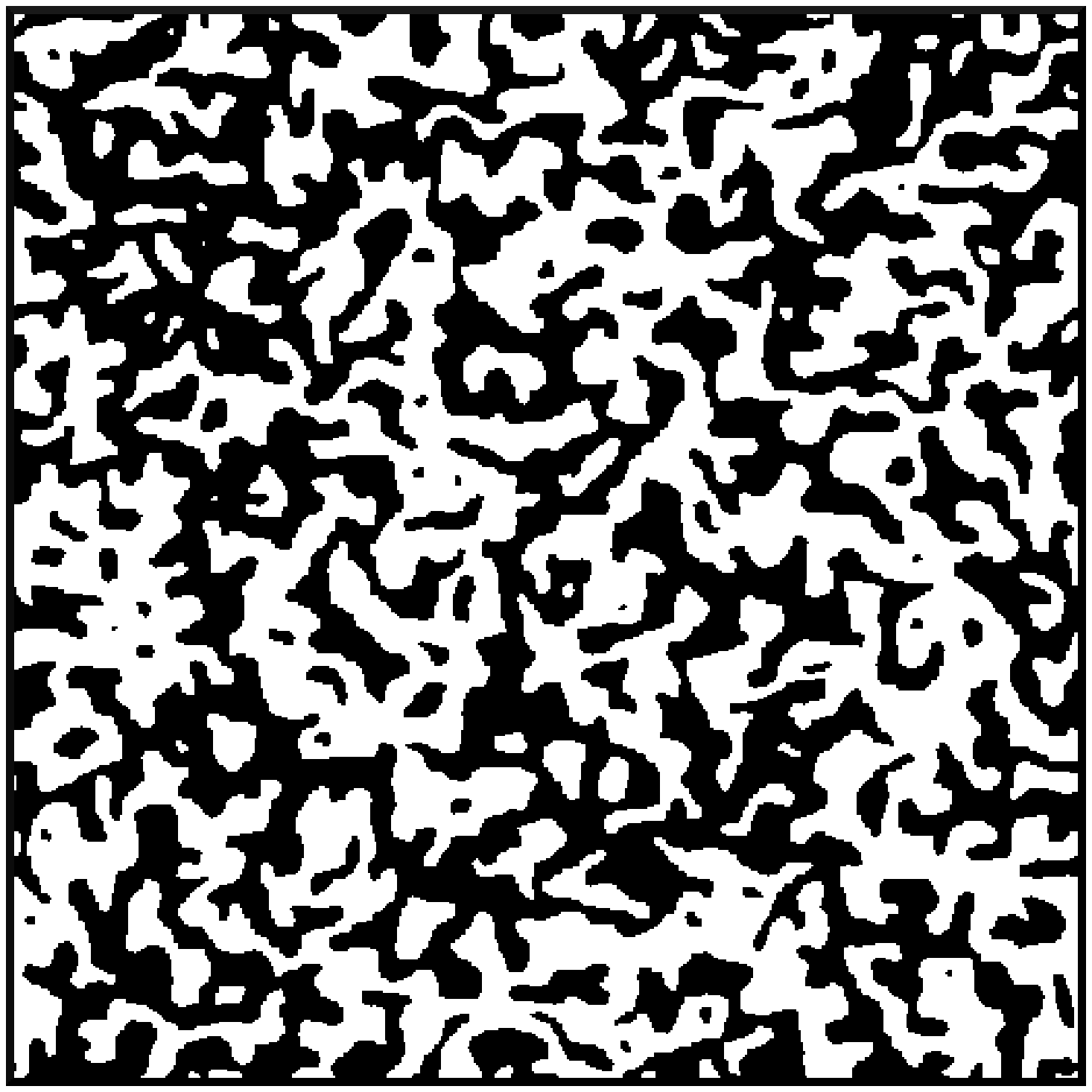}}
\subfigure[]{\includegraphics[width=0.116\textwidth]{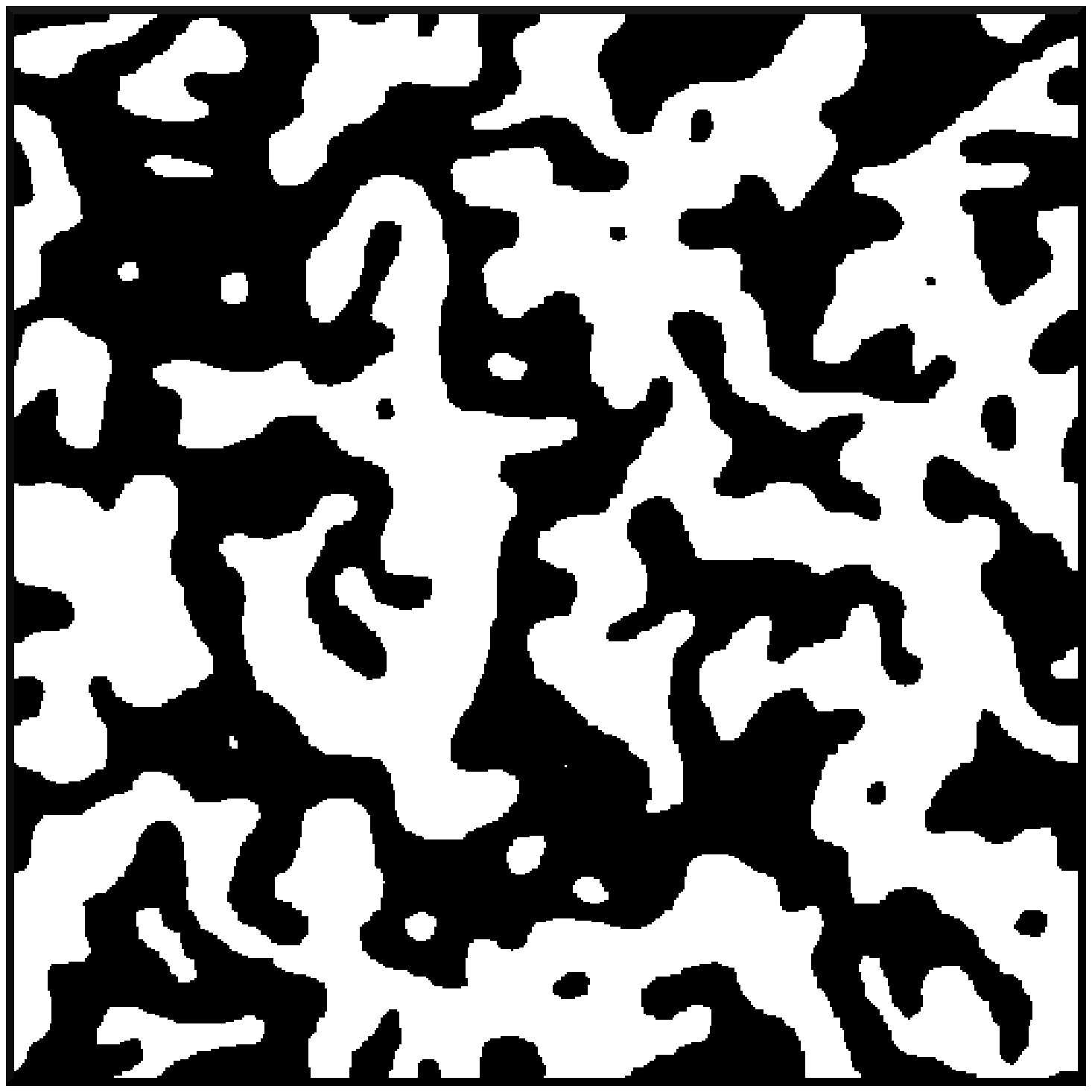}}
\subfigure[]{\includegraphics[width=0.116\textwidth]{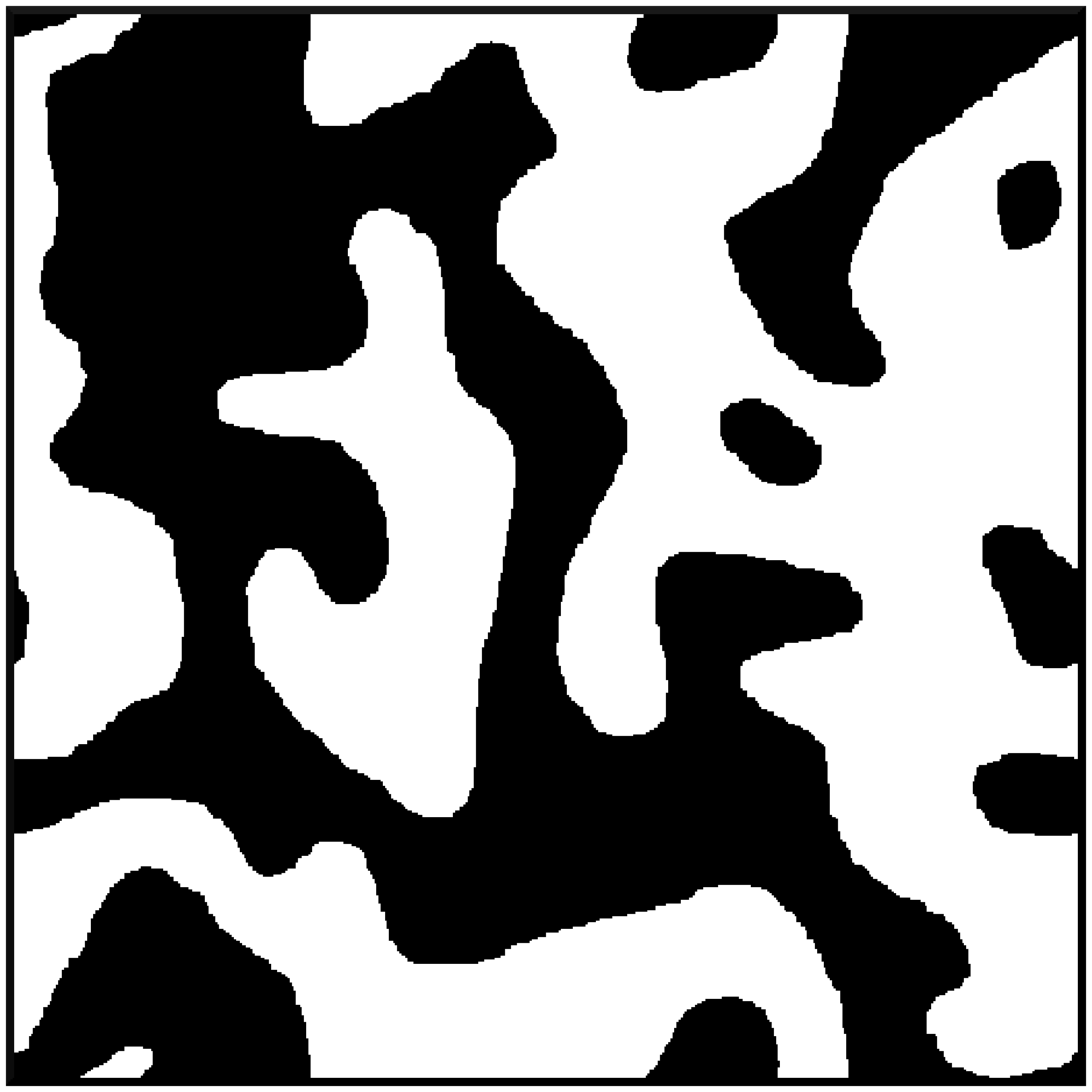}}
\subfigure[]{\includegraphics[width=0.116\textwidth]{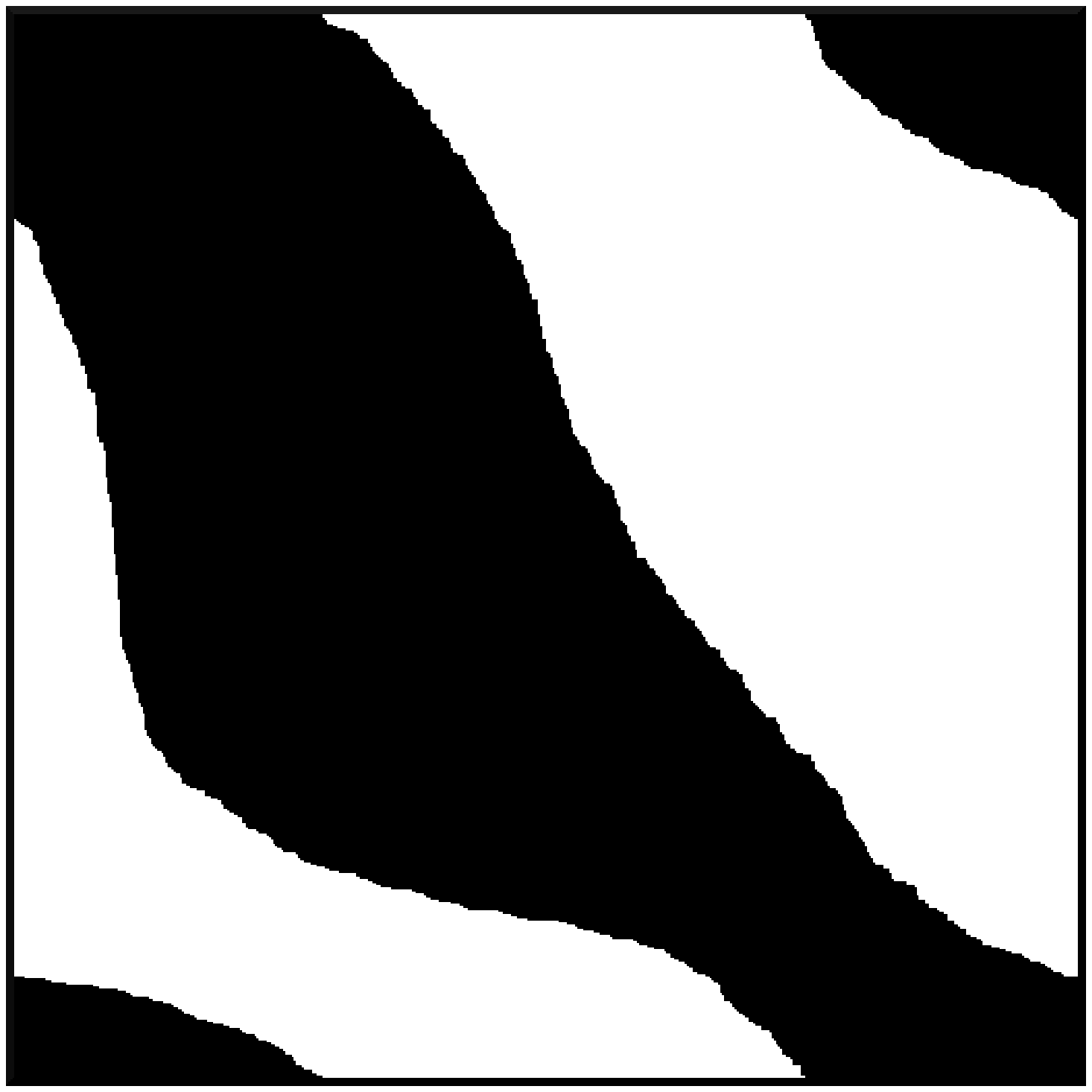}}
\vskip 0.1in
\subfigure[]{\includegraphics[width=0.116\textwidth]{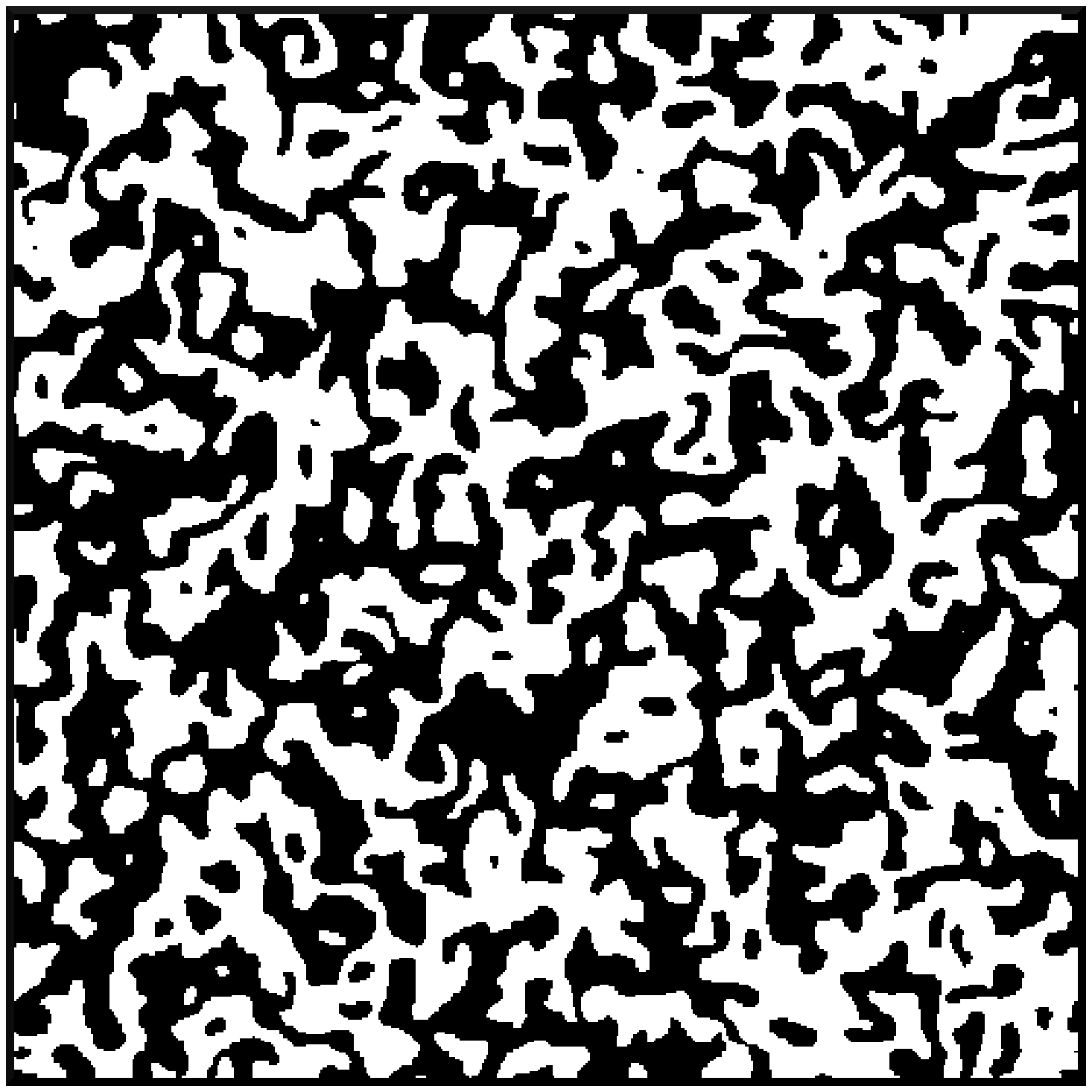}}
\subfigure[]{\includegraphics[width=0.116\textwidth]{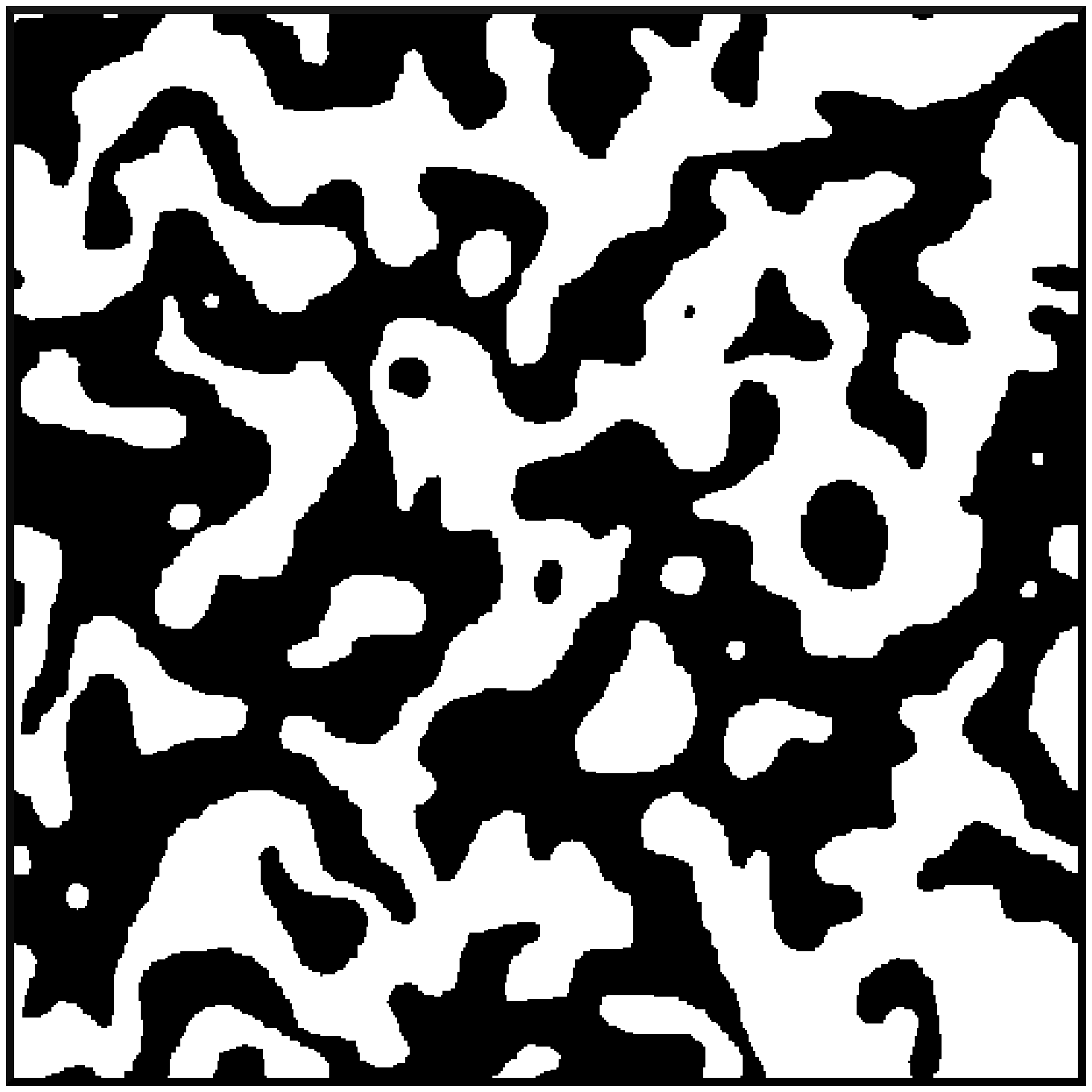}}
\subfigure[]{\includegraphics[width=0.116\textwidth]{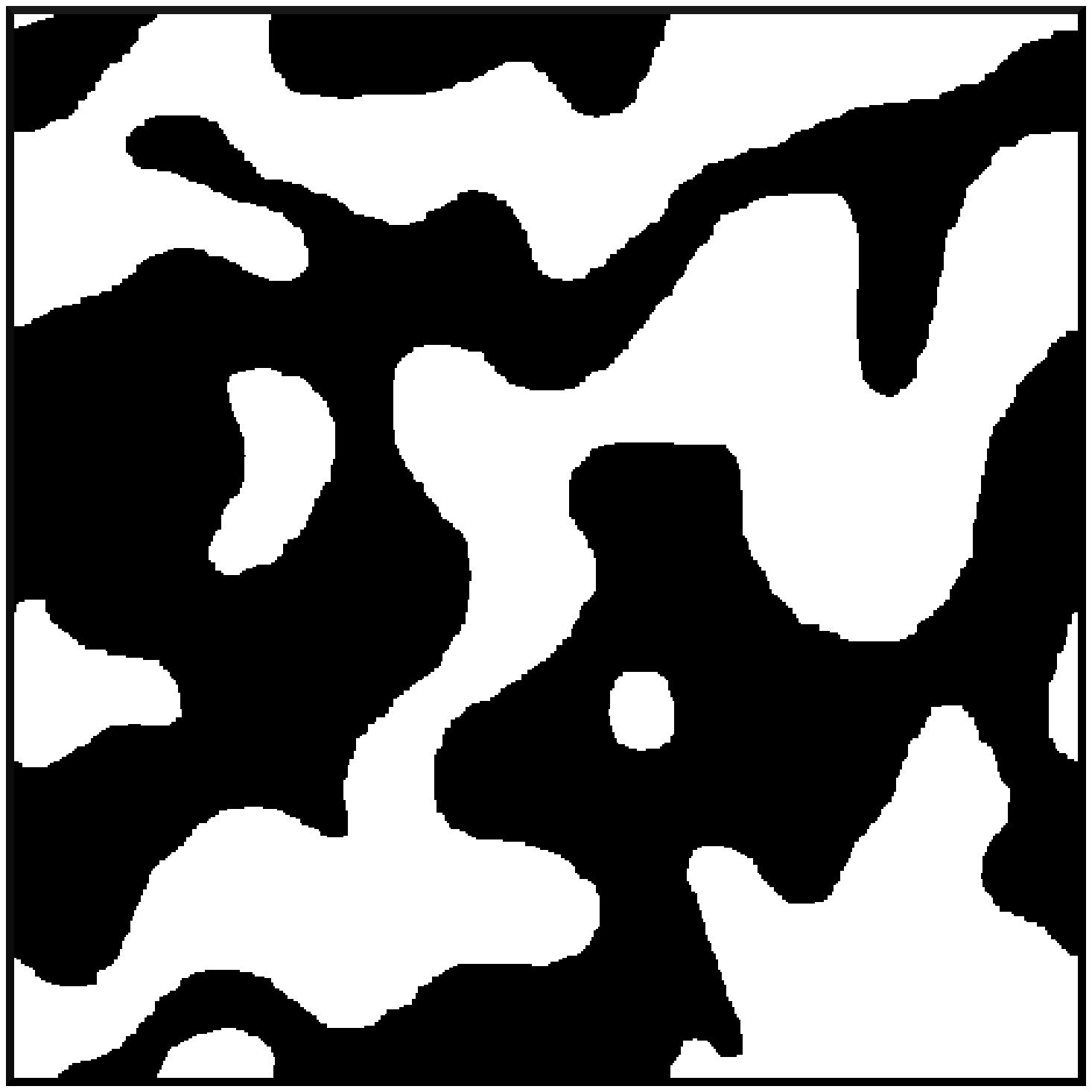}}
\subfigure[]{\includegraphics[width=0.116\textwidth]{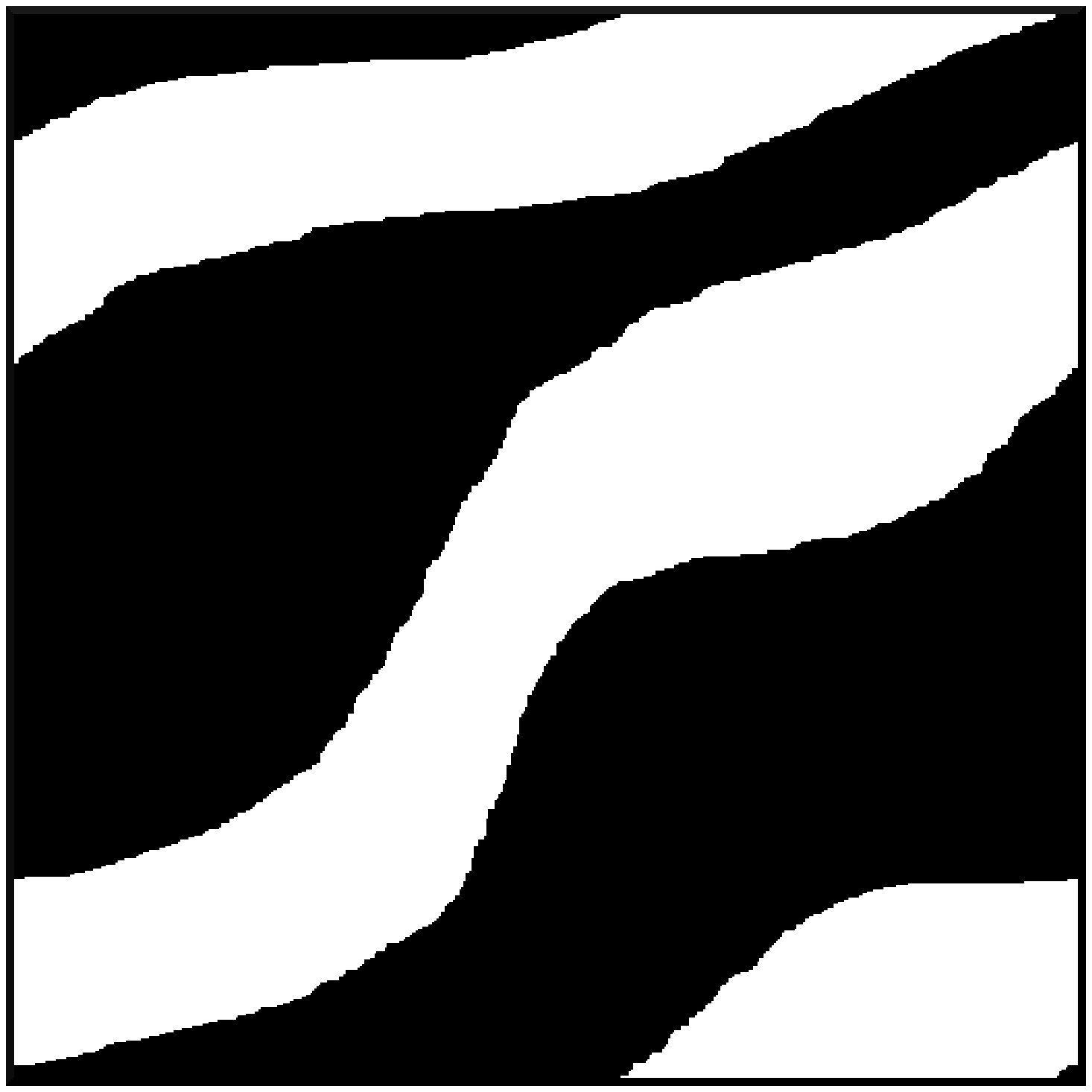}}
\caption{\label{coarsening} Snapshots of coarsening in the nearest-neighbor
  kinetic Ising model on a $1024\times1024$ square lattice with periodic
  boundary conditions at: (a,e) $t=200$, (b,f) 1000, (c,g) 5000, and (d,h)
  50000 after a quench from $T = \infty$ to $T=0$.  Top: evolution to $(1,1)$
  stripes (probability $\approx 0.04$); bottom: evolution to $(2,1)$ stripes
  (probability $\approx 0.00015$). }
\end{figure}

The connection to critical percolation is extraordinarily fruitful because it
allows us to understand why the system may fall into stripe states rather
than ground states and it also predicts the probabilities of various
outcomes~\cite{BKR09}.  For example, the probability to reach a state with
vertical stripes~\cite{caveat} equals the spanning probability
$\mathcal{P}_{0, 1}$ to have a path that spans the system in the vertical
direction at the percolation threshold (and no spanning paths in other
directions).  The spanning probabilities $\mathcal{P}_{0, 1}$ and
$\mathcal{P}_{1,0}$ are exactly
known~\cite{cardy_critical_1992,Pinson_94,Pruessner_04}, and this led to the
prediction that the probability to reach a stripe state equals $0.3390\ldots$
for the square with periodic boundary conditions, in agreement with numerical
simulations~\cite{BKR09}.  (For free boundary conditions this probability is
$\frac{1}{2}-\frac{\sqrt{3}}{2 \pi} \ln\frac{27}{16} = 0.3558\ldots$).

Here we argue that the connection to percolation is much deeper and applies
to a large family of positive-energy metastable states, of which straight
stripes are merely the simplest members.  This connection also applies to a
broad class of coarsening models with non-conserved order-parameter dynamics,
including the time-dependent Ginzburg-Landau equation
(TDGL)~\cite{lif_62,gunton_dynamics_1983,bray_review} and the kinetic Ising
model (KIM).  We will apply the connection to percolation to determine the
probabilities to reach general stripe states that wind $a$ times in one
Cartesian direction and $b$ times in the orthogonal direction for both the
two-dimensional TDGL and KIM with periodic boundary conditions.

The TDGL for a coarse-grained magnetization density $m(\mathbf{r})$ evolves
according to
\begin{equation}
\label{coarse:gl}
\frac{\partial m}{\partial t}=\nabla^2 m-V'(m),
\end{equation}
where $V(m)=\frac{1}{2}(1-m^2)^2$ is the classic double-well potential with
minima at $m=\pm 1$ to account for the equilibrium magnetization of a
ferromagnetic system.  To investigate coarsening that is driven by this TDGL,
we discretize this equation and integrate it forward in time by an explicit
scheme and average results over many zero-magnetization initial conditions.

\begin{center}
\begin{figure}[ht]
\includegraphics[width=0.48\textwidth]{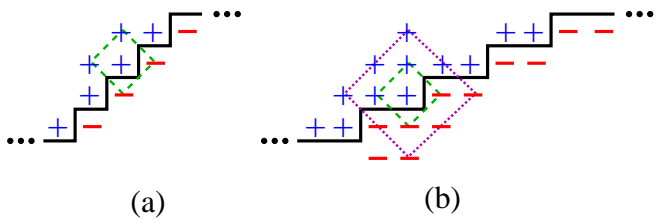}
\caption{(a) $[1,1]^\infty$ staircase interface.  With nearest-neighbor
  interactions (dashed square) interfacial spins can flip freely, but are
  stable with longer-range interactions.  (b) $[2,1]^\infty$ staircase.
  Interfacial spins can flip freely with Manhattan metric first- and
  second-neighbor interactions, but are stable with longer-range
  interactions.}
\label{stairs}
\end{figure}
\end{center}

To reveal the connection to percolation for the discrete KIM, it is essential
to extend this model to more distant interactions.  The Hamiltonian that we
study is
\begin{equation}
\label{H}
\mathcal{H}= -\frac{1}{2}\sum_{i;n} J_n s_i s_{i+n}\,.
\end{equation}
For a given spin $i$, the sum is over the $n^{\rm th}$-nearest neighbors of
$s_i$, where $n^{\rm th}$-nearest neighbor is defined (for convenience) by
the Manhattan metric, in which the distance between $(0,0)$ and $(x,y)$ is
$|x|+|y|$.  We endow this Hamiltonian with single spin-flip
dynamics~\cite{KRB10}.  Operationally, we use Glauber
dynamics~\cite{glauber}; we pick a spin at random and flip it if this event
decreases the energy of the system.  If the energy is unchanged by this flip,
the event is accepted with probability $\frac{1}{2}$.

On the basis of universality~\cite{S71}, cooperative behavior of a
ferromagnet should not fundamentally depend on the interactions as long as
they decay rapidly with distance.  However, there are subtle but important
interaction-range dependent effects that help expose the parallelism between
coarsening in the KIM and critical percolation.  For the KIM with
second-neighbor ferromagnetic interactions of any magnitude, one sees that
the regular $[1,1]^\infty$ staircase shown in Fig.~\ref{stairs} becomes
infinitely long lived.  That is, there is an energy cost to flip any spin on
either side of this staircase.  The stability of this diagonal staircase
causes a stripe state that winds once around a periodic square (a torus) in
both the $x$- and $y$-directions to be infinitely long-lived at zero
temperature.  Similarly, extending the interaction range to third neighbors
additionally causes $[2,1]^\infty$ and $[1,2]^\infty$ staircases to become
infinitely long-lived and thereby stabilize $(2,1)$ and $(1,2)$ stripe states
(Fig.~1(h)).  As the interaction range becomes infinite~\cite{range}, stripe
states with arbitrary integer winding numbers $(a,b)$ are infinitely
long-lived in a square system.

To make the quantitative correspondence between coarsening and percolation,
we need exact results for spanning
probabilities~\cite{cardy_critical_1992,langlands_universality_1992,watts_crossing_1996,smirnov,schramm,maier_2003,dubedat,simmons_percolation_2007},
particularly for the torus topology~\cite{Pinson_94,Pruessner_04}.  As above,
we label spanning clusters by their horizontal and vertical winding numbers,
$a$ and $b$ respectively.  Unique classes of spanning clusters arise for each
pair of values $a,b\ne 0$ in which $a$ and $b$ are co-prime (i.e., $a$ and
$b$ have no common divisors).  Stripes that are characterized by $(a,b)$ and
by $(-a,-b)$ are equivalent and we therefore set $a>0$.

Let $\mathcal{P}_{a,b}(r)$ be the probability for a spanning cluster in
continuum percolation with winding numbers $(a,b)$ on a rectangle with
periodic boundary conditions and with aspect ratio $r\equiv L_y/L_x$.  Here
$L_x$ and $L_y$ are the linear dimensions of the system in the $x$- and
$y$-directions.  For $L_x,L_y\to\infty$, this spanning probability is known
to be~\cite{Pinson_94,Pruessner_04}
\begin{equation}
\label{P_abr}
\mathcal{P}_{a, b}(r) = \frac{\mathcal{Z}_{a, b}(6; r)-2\mathcal{Z}_{a, b}\big(\frac{8}{3};r\big)+\mathcal{Z}_{a, b}\big(\frac{2}{3};r\big)}{2[\eta(e^{-2\pi r})]^2}\,,
\end{equation}
where $\eta(q) = q^{1/24} \prod_{k \geq 1}(1-q^k)$ is the Dedekind $\eta$
function~\cite{AS} and $\mathcal{Z}_{a, b}(G; r)$ is the infinite sum
\begin{equation}
\label{Z}
\mathcal{Z}_{a, b}(G;r) = \sqrt{\frac{G}{r}}\sum_{j=-\infty}^\infty \!\exp\!\left[- \pi G\left(\frac{a^2}{r}+b^2 r\right)\,j^2 \right]\,.
\end{equation}
We tacitly assume that $r\geq 1$; for $r<1$, the spanning probabilities can
be extracted from the obvious duality relation
$\mathcal{P}_{a, b}(r) = \mathcal{P}_{b, a}(\frac{1}{r})$. 

\begin{center}
\begin{figure}[ht]
\includegraphics[width=0.49\textwidth]{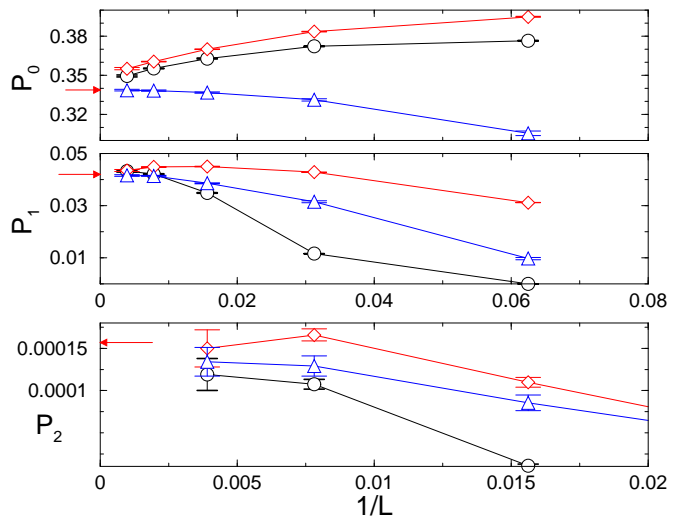}
\caption{(a) The stripe probabilities $P_0$, $P_1$, and $P_2$ versus
  $\frac{1}{L}$ for the KIM with nearest-neighbor ($\circ$) and
  second-neighbor interactions ($\diamond$), and for the TDGL
  ($\bigtriangleup$).  Arrows indicate exact values from Table~I.  For the
  KIM, data are based on $3.2\times 10^6$ realizations for $L\leq 128$ and
  $3.2\times 10^5$ realizations for $L=256$.  For the TDGL, data are based on
  $10^6$ realizations for $L\leq 128$ and $5\times 10^5$ realizations for
  $L=256$.}
\label{probs}
\end{figure}
\end{center}

We study the simplest crossing probabilities for a square $L\times L$ system:
(i) $\mathcal{P}_0=\mathcal{P}_{0,1}+\mathcal{P}_{1,0}= 2\mathcal{P}_{0,1}$,
the probability for a vertical or horizontal stripe, (ii)
$\mathcal{P}_1=\mathcal{P}_{1,1}+\mathcal{P}_{1,-1}= 2\mathcal{P}_{1,1}$, the
probability for a stripe in the $(1,1)$ or $(1,-1)$ directions, and (iii) for
$n\geq 2$, we define $\mathcal{P}_n=4\mathcal{P}_{n,1}$, the probability for
a stripe in the 4 distinct $(\pm n,1)$ and $(\pm 1,n)$ directions.  The
series in Eq.~\eqref{Z} converges rapidly in $j$ and we also make use of the
series representation of the Dedekind $\eta$ function,
\begin{equation*}
\label{eta}
[\eta\big(\rho^{12}\big)]^{-2} = \rho^{-1}(1+2\rho^{12}+5\rho^{24}+10\rho^{36}+\ldots)\,,
\end{equation*}
with $\rho\equiv e^{-\pi/6}$, to give
\begin{equation}
\label{pn}
\begin{split}
  \mathcal{P}_0&= \sqrt{\tfrac{8}{3}}\,\rho^3
  \left(1-\rho^{12}-\rho^{24}+4\rho^{32} + \ldots\right)\\
 \mathcal{P}_1&= \sqrt{\tfrac{8}{3}}\,\rho^7
\left(1 + 2\rho^{12} + 2\rho^{24}+4\rho^{36}+\ldots\right)\\
 \mathcal{P}_n&= \sqrt{\tfrac{32}{3}}\,\rho^{4n^2+3}
\left(1 + 2\rho^{12} + 5\rho^{24}+10\rho^{36}+\ldots\right)\,,
\end{split}
\end{equation}
where the last line holds for all $n\geq 2$.  These stripe probabilities are
given to 4-digit accuracy in Table~I.  Our numerical data for the first three
probabilities (Fig.~\ref{probs}), which are accessible by
simulations, have been obtained by a cluster multilabeling method~\cite{NZ00}.
The extremely good agreement between theory and the simulation results for
both the TDGL and the KIM provides strong evidence that there is indeed an
intimate connection between percolation crossing probabilities and
two-dimensional coarsening.

\begin{table}[ht]
\label{Table_pn}
\begin{tabular}{|c|c|c|c|c|c|}
\hline
$n$                           & 0 & 1 & 2 & 3 & 4 \\
\hline
$\mathcal{P}_{n}$  & 0.3388 & 0.04196 & $1.567\times 10^{-4}$ & $4.438 \times 10^{-9}$ & $1.906 \times 10^{-15}$  \\
\hline
\end{tabular}
\caption{The probabilities $\mathcal{P}_{n}$ for $(n,1)$ stripes on a
  square lattice for small $n$.}
\end{table}

A second natural set of interesting cases are diagonal stripes with tilt
angle $\pm 45^\circ$ on an $L\times nL$ rectangle with periodic boundary
conditions.  Following the same calculational steps as those given previously
for the square system, the series representation for the corresponding
probability $\Pi_n$ is given by
\begin{equation}
\label{D_exp}
\Pi_n  = \sqrt{\tfrac{8}{3n}}\,\rho^{7n}\left[1+2\rho^{12n}+2\rho^{24n}+\ldots  \right]\,,
\end{equation}
where again $\rho\equiv e^{-\pi/6}$.  From this expression, we numerically
obtain the values shown in Table~II (to 4-digit accuracy).  Our simulation
data for $\Pi_n$ for $n=2$ and $n=3$ are consistent with the predictions of
Table~II.  For $n\geq 4$, $\Pi_n$ is so small that is not practical to
accurately measure it by simulations.

\begin{table}[ht]
\label{Table_pin}
\begin{tabular}{|c|c|c|c|c|}
\hline
$n$                            & 1 & 2 & 3 & 4 \\
\hline
$\Pi_n$  &  0.04196 &  $7.567\times 10^{-4}$ &  $1.582\times 10^{-5}$ & $ 3.506\times  10^{-7}$  \\
\hline
\end{tabular}
\caption{$\Pi_n$ for diagonal stripes on a $L\times nL$ rectangle.}
\end{table}

An intriguing feature of arbitrary $(a,b)$ stripe states for the discrete
Ising model is the intricate nature of the staircase interface between
stripes when $a$ and $b$ are both large.  The boundaries between the stripe
states discussed thus far are either perfect straight lines (vertical and
horizontal stripes) or a regular staircase that is inclined at $45^\circ$
(see Fig.~\ref{stairs}(a)).  Stability with respect to single spin-flip
dynamics imposes severe restrictions on the form of these staircases.  For
example, a stripe with winding numbers (1,1) could hypothetically arise from
a regular staircase that consists of alternating vertical and horizontal
steps of length 2.  However, such a staircase is unstable because the energy
is decreased by flipping the corner spins.  This length constraint holds
generally: adjacent vertical and horizontal segments in any stable staircase
cannot both be longer than 2.  Thus the only stable interface for (1,1)
stripes is the regular staircase that we define as ${\bf 1}^\infty$.  This
staircase consists of the periodic sequence of building blocks ${\bf 1}
\equiv [1,1]$, in which $[1,1]$ denotes a unit-length horizontal segment
followed by a unit-length vertical segment.

Continuing this line of reasoning, the only stable staircase in the $(1,n)$
direction is ${\bf n}^\infty$, where ${\bf n}= [1,n]$.  Similarly, $({\bf
  1}{\bf 2})^\infty$ is the stable staircase in the (2,3) direction, $({\bf
  1}{\bf 1}{\bf 2})^\infty$ is the stable staircase in the (3,4) direction,
$({\bf 1}{\bf 2}{\bf 2})^\infty$ is the stable staircase in the (3,5)
direction, etc. The number of staircases going in the same direction is
infinite. For instance, the $({\bf 1}{\bf 1}{\bf 2}{\bf 2})^\infty$ staircase
goes in the (2,3) direction, yet it is unstable. This instability indicates
that there is another general rule to build allowed staircase
interfaces~\cite{KORnew}: only minimal representations are stable.  Analysis
of stable staircases reveals an intriguing connection with the Farey
sequences and the Stern-Brocot tree~\cite{GKP}. To illustrate it, we recall
that for two neighbors in some Farey sequence, e.g. for $\frac{1}{2}$ and
$\frac{1}{3}$, their `sum' is defined via the rule $\frac{1}{2}\oplus
\frac{1}{3} = \frac{1+1}{2+3}=\frac{2}{5}$, and this is taken as an
indication that $({\bf 2}{\bf 3})^\infty$ is the stable staircase in the
(2,5) direction.

The existence of an infinite variety of spanning paths in the KIM with
infinite-range interactions also has intriguing implications for the model
with short-range interactions.  Consider first the classic case of
nearest-neighbor interactions.  A useful diagnostic to detect metastable
stripes with winding numbers $a,b\geq 1$ is to monitor the ``survival
probability'' $S(t)$, defined as the probability that there still exist
flippable spins in the system at time $t$ (the term flippable means that when
such a spin is flipped, the energy of the system either decreases or remains
constant).  If there is a single coarsening time $\tau$ that scales as $L^2$,
then one naturally expects that $S(t)$ should asymptotically decay as
$e^{-t/\tau}$.

\begin{center}
\begin{figure}[ht]
\includegraphics[width=0.4\textwidth]{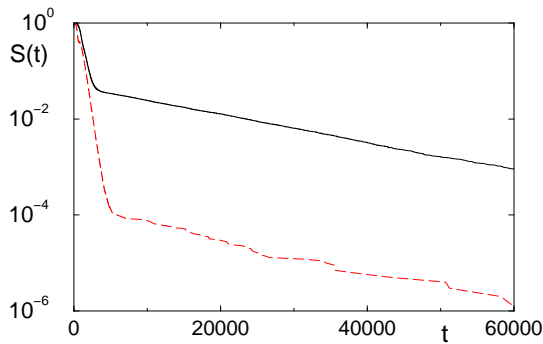}
\caption{ Survival probability $S(t)$ versus $t$ for the KIM on a $64\times
  64$ torus with nearest-neighbor interactions (solid) and second-neighbor
  interactions (dashed).  }
\label{S}
\end{figure}
\end{center}

The actual behavior is markedly different (Fig.~\ref{S}), with the evolution
of $S(t)$ governed by two time scales~\cite{spirin_fate_2001}.  The expected
behavior where $S(t)\sim e^{-t/\tau}$ holds until $S(t)\approx 0.05$.  At
this point, the remaining configurations predominantly have a $(1,1)$ stripe
topology (top line of Fig.~1).  As indicated in Fig.~\ref{stairs}(a) many of
the spins along the interface that separates two diagonal stripes are in
zero-energy environments and can flip with no energy cost.  The fluctuations
of these freely-flippable spins lead to bulk diffusive motion for the
interface.  When two such diffusing interfaces meet, energy-lowering spin
flips occur that ultimately lead the system to the ground state.  The decay
of $S(t)$ in this asymptotic regime is again exponential in time, but now
with characteristic decay time that scales as $L^3$~\cite{spirin_fate_2001}.

For the KIM with (weaker) second-neighbor ferromagnetic interactions, the
vertical and horizontal stripe states, as well as the ${\bf 1}^\infty$
staircase, are all stable at zero temperature in a square system.  Thus at
long times, any remaining metastable states are stripes with still higher
winding numbers.  This feature is reflected in the time dependence of $S(t)$.
The decay of $S(t)$ in the second-neighbor KIM is qualitatively similar to
that of the nearest-neighbor model, but the break in the decay now occurs
when $S(t)\approx 10^{-4}$ (Fig.~\ref{S}).  The long-lived states that remain
beyond this break are predominantly those with winding numbers $(2,1)$ and
$(1,2)$ that ultimately relax to the ground state by interface diffusion.
Such stripe states occur with probability $1.567\times 10^{-4}$ (Table~I),
consistent with the location in the break in the time dependence $S(t)$.
While these types of tilted stripe states are ephemeral when the interaction
range is finite (albeit with a lifetime that grows as $L^3$), they become
permanent when the interaction range becomes long-ranged.

To summarize, we have presented evidence for a close connection between
zero-temperature coarsening of two-dimensional ferromagnets with
arbitrary-range but decaying interactions and critical percolation.  This
connection appears to transcend specific models, as our findings apply
equally well to the time-dependent Ginzburg-Landau equation and to discrete
kinetic Ising models.  The probabilities for either system to evolve to a
state that contains stripe paths with specified winding numbers apparently
coincides with the exactly-known spanning probabilities in two-dimensional
critical percolation.  This equivalence suggests that the domain geometry of
the kinetic ferromagnets coincides with that of continuum percolation at the
critical point.

\begin{acknowledgments}
  \smallskip\noindent We thank Troels R{\o}nnow for an interesting seminar
  question that helped spark this work and helpful manuscript suggestions.
  JO and SR also gratefully acknowledge financial support from NSF grant No.\
  DMR-0906504.
\end{acknowledgments}


\begin{thebibliography}{99}
\bibitem{caveat} For the TDGL, stripes eventually disappaer in a time that
  scales exponentially with system size.  We regard this time as infinite in
  the present context.  For the KIM, straight stripes of widths $\geq 2$ are
  infinitely-long lived.
\bibitem{range} All that is required is that all interactions are
  ferromagnetic and the interactions decay with distance.  Even these
  conditions may be more stringent than necessary.
\end{thebibliography}
\end{document}